\documentclass[11pt,a4paper]{article}
 \usepackage{MyJCAPpub}

 \usepackage{amsmath}
 \usepackage{amsfonts}
 \usepackage{amssymb}

 \allowdisplaybreaks[4]     

 \renewcommand{\bf}{\bfseries}
 \renewcommand{\it}{\itshape}

 \newcommand{\showlabel}[1]{
    \label{#1}}

 \newcommand{\diag}{\rm diag}

 \newcommand{\pd}[2]{\frac{\partial {#1}}{\partial {#2}}}
 
 \newcommand{\pdt}[3]{\frac{\partial^2 {#1}}{\partial {#2} \partial {#3}}}
 
 \newcommand{\nn}{\nonumber}
 \newcommand{\er}[1]{(\ref{#1})}          
 \newcommand{\ol}[1]{\overline{#1}}

 \newcommand{\pb}[2]{                       
    \parbox[t]{#1}{
       \raggedright
       \setlength{\parskip}{1.2ex}          
       #2
    }
 }
 \renewcommand{\L}{Lema\^{\i}tre}
 \newcommand{\LT}{\L-Tolman}
 \newcommand{\Sz}{Szekeres}
 \newcommand{\SC}{Swiss Cheese}

 \renewcommand{\normalfont}{\sffamily}

 \title{A New Type of Exact Arbitrarily Inhomogeneous Cosmology: Evolution of Deceleration in the Flat Homogeneous-On-Average Case}
 
 \author[a]{Charles Hellaby}
 \affiliation[a]{Dept. of Maths. and Applied Maths,
   University of Cape Town,
   Rondebosch,
   7701,
   South Africa}
 \emailAdd{Charles.Hellaby@uct.ac.za}
 
 \date{}

 
 \abstract{ 
   A new method for constructing exact inhomogeneous universes is presented, that allows variation in 3 dimensions.  The resulting spacetime may be statistically uniform on average, or have random, non-repeating variation.  The construction utilises the Darmois junction conditions to join many different component spacetime regions.  In the initial simple example given, the component parts are spatially flat and uniform, but much more general combinations should be possible.  Further inhomogeneity may be added via swiss cheese vacuoles and inhomogeneous metrics.  This model is used to explore the proposal, that observers are located in bound, non-expanding regions, while the universe is actually in the process of becoming void dominated, and thus its average expansion rate is increasing.  The model confirms qualitatively that the faster expanding components come to dominate the average, and that inhomogeneity results in average parameters which evolve differently from those of any one component, but more realistic modelling of the effect will need this construction to be generalised.
 }
 
\begin{document}

 \maketitle

 \section{Motivation}

   Up to now, the options for constructing exact inhomogeneous cosmologies were: (i) a small range of non-homogeneous metrics such as the \LT\ (LT) metric \cite{Lem33,Tol34}, the \Sz (S) metric \cite{Szek75a,Szek75b},%
 \footnote{See \cite{Kra97} for a survey of inhomogeneous cosmologies, \cite{BoKrHeCe09,BoCeKr11} for an overview and review of recent developments, and \cite{Hel09} for a well illustrated quick introduction and discussion of selected recent results.  Also see the CQG issue \cite{CQGIC11} on inhomogeneous cosmologies.}
 their generalisations to non-zero pressure, the {\L} and Szafron metrics \cite{Lem33,Szaf77}, and a number of others with a less believable equation of state (EoS); (ii) the swiss cheese (SC) construction%
 \footnote{For a good list of references see Grenon \& Lake \cite{GreLak10}, footnotes 1 to 4.}
 that inserts spherical structures into a Friedmann-{\L}-Robertson-Walker (FLRW) `background', including multi-level swiss cheese structures; etc.%
 \footnote{The interesting method of Lindquist \& Wheeler \cite{LinWhe57}, has surface layers between the Schwarzschild cells, the nature of which is not clear.  A recent non-vaccum generalisation \cite{CliFer10,Clif10} is an approximate treatment.  In the best current exact multi-black hole metric, there are multiple Reissner-Nordstrom bodies, which have gravitational and electromagnetic `forces' exactly balanced \cite{Pap45,Maj47}, with a $\Lambda$ acceleration added \cite{KasTra93}.}
 We here present a new way to construct exact inhomogeneous cosmologies that are arbitrarily inhomogenous out to all distances.  They don't have an all-enveloping `background' metric, or even an asymptotic one, and they could be inhomogeneous on any scale, but they can also be made to have the same `average' everywhere.  The scale and strength of the inhomogeneities can vary across the spacetime, or be kept statistically similar in all regions, and repeating patterns are also possible.  It should be feasible to generalise this construction to other metrics than those considered here.

   In several recent works, Wiltshire \cite{Wilt07a,Wilt07b,Wilt08,Wilt09,SmaWil10,Wilt11} has discussed a range of unresolved questions about the assumptions underlying the standard approach to cosmological model building.  In attempting to address them, he has propounded some alternative approaches, including the `Cosmological Equivalence Principle', and some deep questioning of what we mean by `averaging', and whether there is a well-defined relationship between an `average' model and real observations.  A key concept, that we attempt to model here, is the idea that void regions expand faster than cluster regions, and that, as time goes by, they occupy an increasing fraction of space, so the `average' expansion rate becomes more and more dominated by the void expansion rate, while observers inhabit regions with little or no expansion, thus generating an apparent acceleration \cite{Buch08,Buch11,Rasa11,KolMarMat09}.  The model below is relatively simple, being a first attempt of its kind, and it does not capture as many aspects of this particular proposal as we hope will be possible with subsequent generalisations.

 \subsection{Inhomogeneity in \LT, \Sz\ and \SC\ Models}

   The {\LT} (LT) metric is spherically symmetric, but radially inhomogeneous, describing a ball of dust particles in comoving coordinates, for which the density and the dynamics depend on both time and radius.  It contains the dust FLRW and Schwarzschild-Kruskal-Szekeres metrics as special cases, and is well suited to describing a black hole in a cosmological background.  It is an excellent first approximation for non-linear gravitational collapse.  As a cosmological model, it is certainly good for putting exact inhomogeneities, with a variety of scales, in an asymptotically uniform spacetime, but strong spherical inhomogeneities at very large radii are not observed.  If one thinks of an LT model as an angular average of an inhomogeneous cosmology that is homogeneous on a sufficiently large scale, then the inhomogeneity should die off with radial distance.  Nevertheless, it has seen extensive use in studying the non-linear evolution of cosmic structures, and in offering an explanation of the dimming of the supernovae.  An asymptotically inhomogeneous model has been proposed to solve the horizon problem.

   The {\Sz} (Sz) metric is even more interesting, since it has no Killing vectors, and thus doesn't suffer from the drawbacks of a high degree of symmetry noted above in the LT models.  It has been used to model voids next to clusters, and even triple structures, and the range of possible structures it can describe is still not known.  Nearly all studies have looked at the quasi-spherical class, which may be thought of as an LT model in which each spherical shell has been displaced relative to the others.  This produces a dipole effect in the shell separation and in the density distribution around each shell.  Thus it may be that this `radial-inhomogeneity-plus-varying-dipole' eventually becomes rather unrealistic at large enough radii.  The little studied quasi-pseudo-spherical class can be thought of as an irregular stacking of hyperboloids (pseudo-spherical shells) that vary in density and evolution, each of which contains an underdensity, or overdensity, due to the `pseudo-dipole'.  The principle inhomogeneity runs from one side of the universe to the other, with an extra variation snaking through the middle.  It does not allow arbitrary inhomogneity in arbirary directions.  Therefore this latter class appears to offer the possibility of very interesting structures and cosmologies, but it hasn't really been explored.

   In swiss cheese models, one starts with an FLRW spacetime, and then cuts spherical holes in it, which may be filled with some other spacetime metric.  For the construction to be valid, the Darmois junction conditions must be satisfied on the boundary between the two spacetimes; each boundary being a timelike 3-surface --- that is, the history of a spherical 2-surface.  Thus the possible interiors are restricted to metrics such as Schwarzschild, {\LT}, Vaidya, {\Sz}.  {\L} and Szafron interiors, that have non-zero pressure, are also possible but have not to our knowledge been used.  Very often, the matching also restricts the FLRW EoS to be that of dust.  Thus there is an obvious `average' FLRW model, and the behaviour of the average is known from the start.

   The new construction has features significantly different from each of these.  Further, it may be combined with all of the above, thereby much expanding the range of possible exact inhomogeneous cosmological models.
 
 \section{Assembling Multiple `Voids' \& `Clusters' in a Single Exact Spacetime}

 \noindent
 ${}$ \hfill
 \pb{14cm}{
 ${}$ \hfill
 \includegraphics{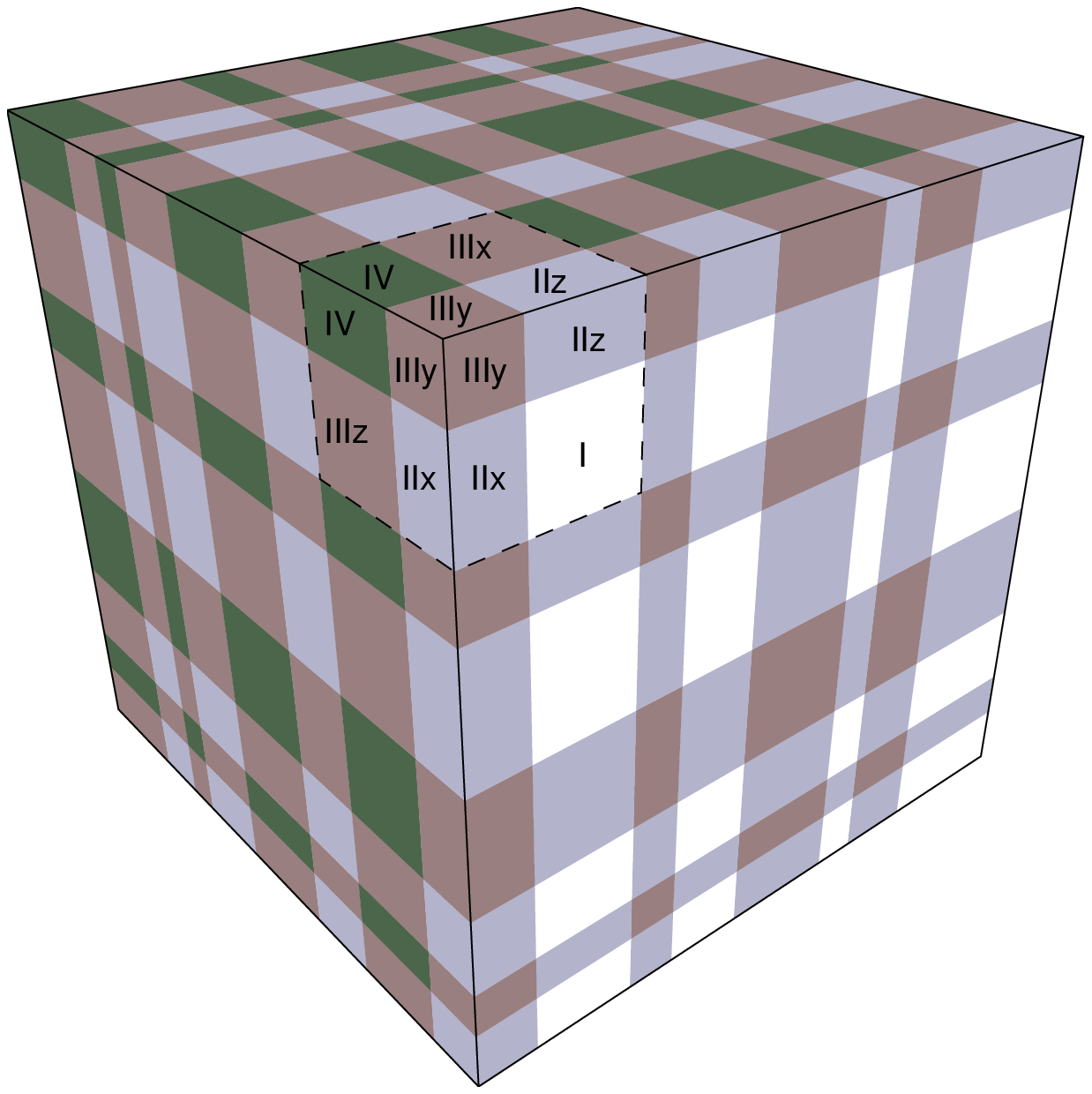}
 \hfill ${}$ \\
 {\small {\bf Figure 1.}  A sample from a multi-component inhomogeneous universe at one moment in time, showing several `repeats' of a `block' of component cuboids.  In this illustration, each block ($4 \times 4 \times 4$ of them shown) has 8 components; 1 of type I, 3 of type II, 3 of type III, and 1 of type IV, and their sizes in each block are actually different, so they are never exactly repeated.  These components are marked on the block at the near corner --- the 3rd type II component (IIy) is of course hidden.  In the main example of the paper, each component is a distinct Kasner-type region.  The different K components have different expansion behaviour, so the fraction of a block they occupy changes with time.  One could construct blocks with $3 \times 3 \times 3$ components, or  $5 \times 2 \times 7$, or even component sequences in each of the $x$, $y$ \& $z$ directions that never repeat.  What matters is that the two expansion rates in the surfaces where two components join are the same in both component regions.  Note that even when the set of expansion rates in a block repeats, the block sizes do not have to.  This is the case illustrated above.}
 }
 \hfill ${}$
 \\[2mm]

   In this section we construct a truly inhomogeneous universe, that does not have an obvious ``background" metric.  We use the Darmois junction conditions in General Relativity (GR) to join many regions of different matter content and evolution type.  If the Darmois conditions are obeyed, the result is an exact solution of the Einstein Field Equations (EFEs).
   
   The essential idea is to do a 3-d tesselation, to fill space with an array of spacetime regions that are properly matched together at their boundaries.  It is the need to ensure a proper 3-surface matching at each surface where component regions join, that renders this otherwise simple idea distinctly non-trivial.  We begin by considering a kind of irregular cubic lattice, in which the basic `unit' or `building block' is composed of 8 `pieces', or `component regions', each one a different Kasner-type (K) metric \cite{Kasn25} --- see Fig.\ 1.  The hope is that this construction may be generalisable to a variety of other metrics.  In fact, an inhomogenous cosmology consisting of multiple slabs of FLRW and Kasner has been constructed in the ``cheese slice" models of \cite{DyeLanSha93,DyeOli01}, but the model presented here is much more realistic in that it is inhomogeneous in all 3 spatial dimensions, instead of just 1.%
 \footnote{Some properties were investigated in \cite{GiaDye09}.  Although \cite{CenMat79} adjoined different Bianchi I cosmologies on an initial time slice, and mentioned the case of FLRW next to Kasner, it is evident that a full Darmois-type matching was not achieved, since ``delta-function discontinuities in the Riemann tensor" are found in section VI - i.e ``surface layers" or ``shock fronts" developed.  A matching of planar dust (Ellis) metrics to vacuum was considered in \cite{Lak92}.}

   Therefore, at any given moment, the 3-d patchwork is like a cubic lattice, but each of the 8 components of a block has different expansion behaviour, so the relative sizes of each component will evolve greatly.  The example illustrated in Fig.\ 1, needs 4 types of region, types I - IV.  We could try to use type I regions to model `voids', type II `walls', type III `filaments`, and type IV `superclusters' (`clusters' for short).  Each type I region adjoins 6 type II regions across a timelike 3-surface; each type IV region adjoins 6 type III regions; each type II region adjoins 4 type III regions and 2 type I; and each type III region adjoins 4 type II regions and 2 type IV.  To make sure the result is a regular solution of the EFEs, we will have to apply the Darmois junction conditions to the loci where 2 or more of these regions meet; see \S \ref{JCs} for the details.
   
   In the simplest case, there is only one kind of type II \& type III region, though oriented differently, and the basic block is cuboidal and repeats indefinitely.  However, as subsequent sections will show, this much regularity is not required by the construction.  In the more general case, the 3 type II \& 3 type III regions may all be different metrics, and the expansion rates and sizes of the blocks and their components need not repeat, but rather have a random distribution of parameters.  See section \ref{MKU}.

 \section{The Kasner-type Metric}

   The Kasner metric\cite{Kasn25}, and its non-vacuum generalisation, is a spatially flat, anisotropic Bianchi model of type I.  It has a different `expansion law' in each of 3 perpendicular directions.  The metric is
 \begin{align}
   ds^2 & = - dt^2 + t^{2 \alpha} \, dx^2 + t^{2 \beta} \, dy^2 + t^{2 \gamma} \, dz^2   \showlabel{Kds2} ~.
 \end{align}
 The Einstein field equations (EFEs), and the expansion are given by
 \begin{align}
   \kappa \rho & = \frac{\alpha \beta + \beta \gamma + \gamma \alpha}{t^2} ~, \\
   \kappa p_x & = \frac{\beta + \gamma - (\beta^2 + \beta \gamma + \gamma^2)}{t^2} ~, \\
   \kappa p_y & = \frac{\gamma + \alpha - (\gamma^2 + \gamma \alpha + \alpha^2)}{t^2} ~, \\
   \kappa p_z & = \frac{\alpha + \beta - (\alpha^2 + \alpha \beta + \beta^2)}{t^2} ~, \\
   \Theta & = \frac{\alpha + \beta + \gamma}{t} ~.
 \end{align}
 The pressures are all proportional to the density, but different in the 3 perpendicular directions.
 The unit vector $u^a = \delta^a_t$ is geodesic and comoving with the matter.

 \paragraph{The Minkowski Case}

   To make the Riemann tensor zero requires e.g.
 \begin{align}
   \alpha = 0 = \beta = \gamma ~;~~~~~~\mbox{or}~~~~~~ \alpha = 0 = \beta ~,~ \gamma = 1 ~.
 \end{align}

 \paragraph{The Vacuum Case}

   The requirement $\rho = 0 = p_x = p_y = p_z$ leads to $(\alpha, \beta, \gamma)$ being an equally spaced triplet round the Kasner circle:
 \begin{align}
   \alpha & = \frac{1 + 2 \cos\theta}{3} ~, \\
   \beta & = \frac{1 + 2 \cos(\theta + 2 \pi/3)}{3} ~, \\
   \gamma & = \frac{1 + 2 \cos(\theta + 4 \pi/3)}{3} ~.
 \end{align}
 It is not possible for more than two of $\alpha$, $\beta$ \& $\gamma$ to be the same, so vacuum Kasner cannot be isotropic, unless it is Minkowski.  The only vacuum case with two of them the same is $\{\alpha, \beta, \gamma\} = \{2/3, 2/3, -1/3\}$.

 \paragraph{The Isotropic Case}

   Putting $\gamma = \beta = \alpha$ gives the spatially flat FLRW models,
 \begin{align}
   S = t^\alpha ~,~~~~~~
   \kappa \rho = \frac{3 \alpha^2}{t^2} ~,~~~~~~
   \kappa p = \frac{2 \alpha - 3 \alpha^2}{t^2} = \left( \frac{2}{3 \alpha} - 1 \right) \rho ~,
 \end{align}
 where $S$ is the FLRW scale factor.

 \paragraph{The Zero Pressure Case}

   The requirement $0 = p_x = p_y = p_z$ is satisfied by the Minkowski case, the set of vacuum cases, and the $\alpha = 2/3$ FLRW case.

 \subsection{Physicality Conditions}
 \showlabel{PhysCond}

   Since we will be assembling many different K regions to construct an inhomogeneous universe, it is useful to check their physical behaviour.  The condition for non-negative density is easy to satisfy,
 \begin{align}
   \rho \geq 0 & ~~~~\to~~~~ \alpha \beta + \beta \gamma + \gamma \alpha \geq 0 ~,
 \end{align}
 and that for non-negative pressure,
 \begin{align}
   p_x \geq 0 & ~~~~\to~~~~ \beta + \gamma - (\beta^2 + \beta \gamma + \gamma^2) \geq 0 ~, \nn \\
   & ~~~~\to~~~~ \frac{1}{2} \left( (1 - \beta) - \sqrt{(1 - \beta)(1 + 3\beta)}\; \right) \leq \gamma \leq \frac{1}{2} \left( (1 - \beta) + \sqrt{(1 - \beta)(1 + 3\beta)}\; \right) ~, \nn \\
   & ~~~~~~~~~~~ -1/3 \leq \beta \leq 1 ~~~~~~ \big[ -1/3 \leq \gamma \leq 1 \big] ~, \\
   p_y \geq 0 & ~~~~\to~~~~ \gamma + \alpha - (\gamma^2 + \gamma \alpha + \alpha^2) \geq 0 ~, \\
   p_z \geq 0 & ~~~~\to~~~~ \alpha + \beta - (\alpha^2 + \alpha \beta + \beta^2) \geq 0 ~,
 \end{align}
 which is a fattened region around the Kasner circle, with a 3-lobed shape, a bit like a cardamom pod, is also not hard to satisfy.  For a non-relativistic gas, we expect,
 \begin{align}
   p_x < \frac{\rho}{3} & ~~~~\to~~~~ 3(\beta + \gamma) - 3(\beta^2 + \beta \gamma + \gamma^2) < \alpha \beta + \beta \gamma + \gamma \alpha \nn \\
   & ~~~~\to~~~~ \gamma < \frac{1}{6} \left( 3 - 4 \alpha - \beta + \sqrt{9 + 12 \alpha - 6 \beta - 20 \alpha^2 - 4 \alpha \beta + \beta^2}\; \right) \nn \\
   & ~~~~~~\mbox{\&}~~~~~ \gamma > \frac{1}{6} \left( 3 - 4 \alpha - \beta - \sqrt{9 + 12 \alpha - 6 \beta - 20 \alpha^2 - 4 \alpha \beta + \beta^2}\; \right) ~, \\
   p_y < \frac{\rho}{3} & ~~~~\to~~~~ 3(\gamma + \alpha) - 3(\gamma^2 + \gamma \alpha + \alpha^2) < \alpha \beta + \beta \gamma + \gamma \alpha \nn \\
   & ~~~~\to~~~~ \gamma < \frac{1}{6} \left( 3 - 4 \beta - \alpha + \sqrt{9 + 12 \beta - 6 \alpha - 20 \beta^2 - 4 \beta \alpha + \alpha^2}\; \right) \nn \\
   & ~~~~~~\mbox{\&}~~~~~ \gamma > \frac{1}{6} \left( 3 - 4 \beta - \alpha - \sqrt{9 + 12 \beta - 6 \alpha - 20 \beta^2 - 4 \beta \alpha + \alpha^2}\; \right) ~, \\
   p_z < \frac{\rho}{3} & ~~~~\to~~~~ 3(\alpha + \beta) - 3(\alpha^2 + \alpha \beta + \beta^2) < \alpha \beta + \beta \gamma + \gamma \alpha \nn \\
   & ~~~~\to~~~~ \gamma < 3 - 3 \alpha - \beta - \frac{2 \beta^2}{\alpha + \beta} ~.
 \end{align}
 Similarly, to ensure the sound speed less is than light speed, $p < \rho$, we obtain a similar set of conditions.  
 These are not always so easy to obey, because, in the transition from dust FLRW, $(\alpha,\beta,\gamma) = (2/3, 2/3, 2/3)$ to Minkowski vacuum, $(\alpha,\beta,\gamma) = (0, 0, 0)$, the matter becomes increasingly stiff.  Thus in the general case we have an exotic fluid.  Nevertheless, as a first simple example of this type of construction, the above behaviour is not too bad.  In any case, these component regions may themselves be thought of as averages over a more complicated matter distribution, so the above is only an effective bulk EoS (equation of state).  The fact that the pressures are different in different directions is only to be expected in regions with ``pancake" or ``cigar" expansion.  More general metrics for the component spacetimes that are stitched together in such a patchwork, will no doubt allow us to improve this aspect.

 \subsection{Cosmological Units}
 \showlabel{CGU}

   We choose geometric units such that $G = 1 = c$, and for the remaining freedom, we specify that $1$ time unit = $10$~gigayears.  We call these cosmological geometric units, and the values of the cosmological time, length and distance units are:
 \[
   \begin{tabular}{llll}
     1 ctu & = & 10 Gy \\
     1 clu & = & 3.066 Gpc \\
     1 cmu & = & 6.409 $\times 10^{22}$ M$_\odot$
   \end{tabular}
 \]
 In these units, $100$~km/s/Mpc is very close to $1$/ctu.

 \section{Junction Conditions}
 \showlabel{JCs}

   We now implement the Darmois \cite{Darm27} junction conditions.  If they are satisfied, the combined spacetime metric is $C^1$ and piecewise $C^3$, and it may be shown \cite{HelDra94}%
 \footnote{Although this paper is about conservation failing at a signature change surface, the non-signature-changing case was done first.}
 that, due to the Israel identities \cite{Isra66}, the conservation laws $\nabla_\nu G^{\mu\nu} = 0$ are satisfied even through the $C^1$ junctions.

   For the 3-surface $x = X =$~const, we define the surface $\Sigma$ and the surface coordinates $\xi^i$ to be
 \begin{align}
   x_\Sigma^\mu & = (t, X, y, z) ~, \\
   \xi^i & = (t, y, z) ~.
 \end{align}
 Here, greek indices range $0$ to $3$, and latin indices range $1$ to $3$, but note that, in the context of junction conditions, index $0$ indicates the direction orthogonal to $\Sigma$, and is not necessarily time.  We then calculate, in order, the surface basis vectors, their derivatives, the surface normal, the intrinsic metric and the extrinsic curvature:
 \begin{align}
   (e_i)^\mu & = \pd{x^\mu}{\xi^i} = 
      \begin{pmatrix}
         1 & 0 & 0 & 0 \\
         0 & 0 & 1 & 0 \\
         0 & 0 & 0 & 1
      \end{pmatrix} ~, \\
   \pdt{x^\mu}{\xi^i}{\xi^j} & = 0 ~, \\
   n_\mu & = (0, t^\alpha, 0, 0) ~, \\
   {}^3\!\!g_{ij} & = g_{\mu\nu} \, (e_i)^\mu \, (e_j)^\nu = \diag(-1, t^{2\beta}, t^{2\gamma}) ~, \\
   K_{ij} & = - n_\sigma \left( \pdt{x^\sigma}{\xi^i}{\xi^j} + \Gamma^\sigma{}_{\mu\nu} \, (e_i)^\mu \, (e_j)^\nu \right) = 0 ~.
 \end{align}

   The two manifolds we wish to join are labelled ``$+$" and ``$-$", and are both Kasner-type (K) spacetimes:
 \begin{align}
   V^+ :~~~~ x_+^\mu = (t_+, x_+, y_+, z_+) ~,~~~~ g_{\mu\nu} = \diag(-1, t_+^{\alpha_+}, t_+^{\beta_+}, t_+^{\gamma_+}) ~, \\
   V^- :~~~~ x_-^\mu = (t_-, x_-, y_-, z_-) ~,~~~~ g_{\mu\nu} = \diag(-1, t_-^{\alpha_-}, t_-^{\beta_-}, t_-^{\gamma_-}) ~.
 \end{align}
 If the two surfaces to be identified, $\Sigma_+$ \& $\Sigma_-$, are located at $x = X_+$ \& $x = X_-$, and the mapping between them is
 \begin{align}
   t_+ & = t_- = \xi^1 ~, \\
   y_+ & = y_- = \xi^2 ~, \\
   z_+ & = z_- = \xi^3 ~,
 \end{align}
 then the Darmois conditions, which use square brackets to denote the jump in a quantity across the junction, consequently require
 \begin{align}
   \big[ {}^3\!\!g_{ij} \big] = {}^3\!\!g_{ij} |_\Sigma^+ & - {}^3\!\!g_{ij} |_\Sigma^- = 0 ~, \\
   \big[ K_{ij} \big] = K_{ij} |_\Sigma^+ & - K_{ij} |_\Sigma^- = 0 ~.
 \end{align}
 For our matching, these lead to
 \begin{align}
   \beta_+ = \beta_- ~,~~~~~~ \gamma_+ = \gamma_- ~.
 \end{align}
 This is consistent with the Israel requirement \cite{Isra66}, which means that, for an observer moving with a timelike 3-surface, the pressures must match, but the density does not have to.%
 \footnote{At the level of bulk fluid parameters, the Darmois matching is complete.  If however one introduces a kinetic theory description of the matter --- far too complex for present purposes --- then there would be further conditions needed to match all the modes.}  
 Therefore, two K spaces may be joined across any pair of constant $x$ surfaces if the two expansion indices in the plane of the surfaces are the same on either side of the junction.  The expansion index perpendicular to $\Sigma$ may differ in $V_+$ \& $V_-$.  The same goes for matching pairs of constant $y$ or $z$ surfaces.  Thin boundaries where the tangential pressures are different on either side are of course unrealistic; at the atomic level, one would expect streaming of particles and photons to blur the boundaries.  However, junction condition methods are understood to be useful approximations to transitions that happen within a relatively thin region, for which the exact smooth equations are unsolvable.

   Now the Darmois conditions allow the regular junction of two manifolds at a pair of identified 3-surfaces, the 3-surfaces being unbounded.  In our case, each matching surface is bounded by other matching surfaces, so the size of the two identified surfaces must also be the same.  Where 2 cuboidal regions meet, in our array of different K regions, they must have 2 of the coordinate dimensions, $\Delta x$, $\Delta y$ \& $\Delta z$, in common --- see below.

   Furthermore, where 4 cuboids touch at a 2-surface (a 1-d curve existing through time), the corner conditions \cite{Tayl04} should be checked.%
 \footnote{See \cite{GiaDye09} for an application.}  
 However, we expect no difficulty, as the spatial coordinates in the K metric are Cartesian.  Similarly, at the worldlines of the vertices  where 8 cuboids meet, an appropriate modification of those corner conditions should be applied.

 \section{Multi-Kasner Universes}
 \showlabel{MKU}

   We are now ready to patch together Kasner-type (K) cuboids.  Using the matching rules just obtained, a general, repeating $2 \times 2 \times 2$ model would be as in Table 1:
 \\[2mm]
 \noindent
 ${}$ \hfill
 \pb{14cm}{
 ${}$ \hfill
 \begin{tabular}{l|l|l|l}
 Region & Region type & $(\alpha, \beta, \gamma)$ & $(\Delta x, \Delta y, \Delta z)$ \\
  \hline
 I & `void' & $(a, b, c)$ & $(A, B, C)$ \\
 IIx & $x$ `wall' & $(d, b, c)$ & $(D, B, C)$ \\
 IIy & $y$ `wall' & $(a, e, c)$ & $(A, E, C)$ \\
 IIz & $z$ `wall' & $(a, b, f)$ & $(A, B, F)$ \\
 IIIx & $y$-$z$ `filament' & $(a, e, f)$ & $(A, E, F)$ \\
 IIIy & $z$-$x$ `filament' & $(d, b, f)$ & $(D, B, F)$ \\
 IIIz & $x$-$y$ `filament' & $(d, e, c)$ & $(D, E, C)$ \\
 IV & `cluster' & $(d, e, f)$ & $(D, E, F)$
 \end{tabular}
 \hfill ${}$ \\[1mm]
 {\small {\bf Table 1.}  A repeating $2 \times 2 \times 2$ multi-Kasner model.  The region numbers correspond to those marked in Fig. 1.  The triplet $(\alpha, \beta, \gamma)$ is the set of exponents in the metric \er{Kds2}, and the particular values $a$, $b$, $c$, $d$, $e$, $f$ may be chosen freely, subject to appropriate physicality conditions, as in section \ref{PhysCond}.  Similarly the triplet $(\Delta x, \Delta y, \Delta z)$ gives the three coordinate dimensions of a region, as at the end of section \ref{JCs}, and $A$, $B$, $C$, $D$, $E$, $F$ may be chosen freely.}
 }
 \hfill ${}$
 \\[2mm]
 For a simple model, one may set
 \begin{align}
   c = b = a ~,~~~~ f = e = d ~,~~~~
   C = B = A ~~~~\mbox{and}~~~~ F = E = D ~.
   \showlabel{SimpEx}
 \end{align}
 Thus, although this is not necessary, one of the 8 components could easily be an isotropic (i.e. FLRW) region, and we could even have two distinct FLRW regions per block.

   In fact, the building blocks do not have to repeat exactly.  Referring again to Fig. 1, as we follow a line of component regions in the $x$ direction, say, the particular pairs $(\beta, \gamma)$ and $(B, C)$ must be the same for every region along that line, but need not be the same as adjacent lines.  Similarly, following a line in the $y$ direction, the particular pairs $(\alpha, \gamma)$ and $(A, C)$ must be the same all down that row; etc. for following a $z$ line.  Thus, in each of the $x$, $y$ \& $z$ directions respectively, we may have non-repeating sequences of expansion rates and widths: 
$(\alpha_1,\;A_1),\;(\alpha_2,\;A_2),\;(\alpha_3,\;A_3),\; \cdots$; 
$(\beta_1,\;B_1),\;(\beta_2,\;B_2),\;(\beta_3,\;B_3),\; \cdots$; 
$(\gamma_1,\;C_1),\;(\gamma_2,\;C_2),\;(\gamma_3,\;C_3),\; \cdots$.  
Each $(\alpha_i,\;A_i)$ stays constant in a constant $(y, z)$ surface, and so on.  Thereby we may create an arbitrarily inhomogeneous universe.  If desired, these parameter values may be chosen from some statistical distribution.

   Furthermore, if there are FLRW regions, each one of them may be filled with a variety of swiss-cheese inhomogeneities, or even multi-level swiss-cheese inhomogeneites, and exact inhomogeneous models, such as the {\LT} and {\Sz} metrics, can be used in the SC constructions.

 \section{Averaging}

   We now consider how to choose an FLRW model that best approximates this universe in some kind of `average' sense, and we look at the evolution of the `average' Hubble, deceleration and EoS parameters.

   The simplest approach is to do a volume-weighted average.  When the 3-spaces are flat, this is quite simple.  For simplicity, we consider a repeating $2 \times 2 \times 2$ block, so that there is a uniform large-scale average.  The volume occupied by each of the 8 component regions is just
 \begin{align}
   V = \Delta x \, \Delta y \, \Delta z \, t^{\alpha+\beta+\gamma} ~,
 \end{align}
 and its derivatives are
 \begin{align}
   \dot{V} & = \Delta x \, \Delta y \, \Delta z \, (\alpha + \beta + \gamma) \, t^{\alpha+\beta+\gamma-1} ~, \\
   \ddot{V} & = \Delta x \, \Delta y \, \Delta z \, (\alpha + \beta + \gamma)(\alpha + \beta + \gamma - 1) \, t^{\alpha+\beta+\gamma-2} ~.
 \end{align}
 Thus the volume fraction of any given region $i$ is 
 \begin{align}
   f_i = \frac{V_i}{\sum_{j=1}^8 V_j} ~,
 \end{align}
 the volume-averaged expansion rate is 
 \begin{align}
   \ol{\Theta} = \frac{\sum_{i=1}^8 \dot{V}_i}{\sum_{j=1}^8 V_j} = 3 \ol{H} ~,
 \end{align}
 and the volume-averaged deceleration parameter is 
 \begin{align}
   \ol{q} = 2 - \frac{3 \big( \sum_{j=1}^8 V_j \big) \big( \sum_{k=1}^8 \ddot{V}_k \big) }{ \big( \sum_{i=1}^8 \dot{V}_i \big)^2 } ~.
 \end{align}
   If we try to fit an FLRW model
 \begin{align}
   \kappa \rho = \frac{3 (\dot{S}^2 + k)}{S^2} = 3 H^2 (1 + \Omega_k) ~,~~~~~~ \kappa p = \frac{( 2 S \ddot{S} + \dot{S}^2 + k)}{S^2} = H^2 (1 - 2 q + \Omega_k) ~,
 \end{align}
 to this inhomogeneous cosmology, then we would adjust the EoS, $p = w \rho$, so as to reproduce the actual expansion, i.e.
 \begin{align}
   w & = \frac{1}{3} \left( 1 - \frac{2 q}{1 + \Omega_k} \right) ~,
 \end{align}
 and setting $k = 0$ (since the K component regions are all flat), we have the effective EoS parameter
 \begin{align}
   \ol{w} & = \frac{1 - 2 \ol{q}}{3} ~.
 \end{align}

   For the simple example \er{SimpEx} above, these would be,
 \begin{align}
   f_i & = \frac{(t^{3a},~ t^{2a+d},~ t^{a+2d},~ t^{3d})}{(t^a + t^d)^3} ~, \\
   \ol{\Theta} & = \frac{3 (a t^a + d t^d)}{t (t^a + t^d)} ~, \\
   \ol{q} & = - \frac{(t^a + t^d) \big( a (a - 1) t^a + d (d - 1) t^d \big) }{(a t^a + d t^d)^2} ~;
 \end{align}
 and for a single K component they are
 \begin{align}
   f = 1 ~,~~~~~~ \Theta = \frac{\alpha + \beta + \gamma}{t} ~,~~~~~~ q = - 1 + \frac{3}{\alpha + \beta + \gamma} ~,~~~~~~ w = 1 - \frac{2}{\alpha + \beta + \gamma} ~.   \showlabel{fThqw}
 \end{align}

 \section{Model Details and Results}
 
   We now present 3 variations of a specific case, a repeating $2 \times 2 \times 2$ block.  The details of models (A) to (C) are given in tables 2 to 4.  In each, the type I \& type IV regions are specified, and the remaining regions are fixed by the junction conditions.  The repeating pattern allows an average to be calculated that is globally uniform, and thus the evolution of the component parts can be compared with the average.  If the comoving size of a component region is $A \times B \times C$ in the $x$, $y$ \& $z$ directions, then the $x$ size of the region today ($t \sim 1$~ctu) is $\sim A$~clu; so $A = 0.02$ corresponds to $\sim 60$Mpc --- the scale of voids, walls, filaments, etc.  $B$ \& $C$ would have a similar order of magnitude.
 \\[2mm]
 \noindent
 ${}$ \hfill
 \pb{14cm}{
 ${}$ \hfill
 \begin{tabular}{l|l|l|l|l|l|l}
 Region & Region type & Expansion rate triplet $\alpha$-$\beta$-$\gamma$ & $\rho$ & $(p_x, p_y, p_z)$ & $\Theta$ & $q$ \\
  \hline
  &&&&&\\[-3.5mm]
 I & `void' & $(1/2, 1/2, 1/2)$ & $\frac{3}{4 t^2}$ & $(\frac{1}{4 t^2}, \frac{1}{4 t^2}, \frac{1}{4 t^2})$ & $\frac{3}{2 t}$ & $1$ \\[2mm]
 IIx & $y$-$z$ `wall' & $(2/3, 1/2, 1/2)$ & $\frac{11}{12 t^2}$ & $(\frac{1}{4 t^2}, \frac{5}{36 t^2}, \frac{5}{36 t^2})$ & $\frac{5}{3 t}$ & $\frac{4}{5}$ \\[2mm]
 IIy & $z$-$x$ `wall' & $(1/2, 2/3, 1/2)$ & $\frac{11}{12 t^2}$ & $(\frac{5}{36 t^2}, \frac{1}{4 t^2}, \frac{5}{36 t^2})$ & $\frac{5}{3 t}$ & $\frac{4}{5}$ \\[2mm]
 IIz & $x$-$y$ `wall' & $(1/2, 1/2, 2/3)$ & $\frac{11}{12 t^2}$ & $(\frac{5}{36 t^2}, \frac{5}{36 t^2}, \frac{1}{4 t^2})$ & $\frac{5}{3 t}$ & $\frac{4}{5}$ \\[2mm]
 IIIx & $x$ `filament' & $(1/2, 2/3, 2/3)$ & $\frac{10}{9 t^2}$ & $(0, \frac{5}{36 t^2}, \frac{5}{36 t^2})$ & $\frac{11}{6 t}$ & $\frac{7}{11}$ \\[2mm]
 IIIy & $y$ `filament' & $(2/3, 1/2, 2/3)$ & $\frac{10}{9 t^2}$ & $(\frac{5}{36 t^2}, 0, \frac{5}{36 t^2})$ & $\frac{11}{6 t}$ & $\frac{7}{11}$ \\[2mm]
 IIiz & $z$ `filament' & $(2/3, 2/3, 1/2)$ & $\frac{10}{9 t^2}$ & $(\frac{5}{36 t^2}, \frac{5}{36 t^2}, 0)$ & $\frac{11}{6 t}$ & $\frac{7}{11}$ \\[2mm]
 IV & `cluster' & $(2/3, 2/3, 2/3)$ & $\frac{4}{3 t^2}$ & $(0, 0, 0)$ & $\frac{2}{t}$ & $\frac{1}{2}$
 \end{tabular}
 \hfill ${}$ \\[1mm]
 {\small {\bf Table 2.}  Model (A).  Type I regions (voids) are radiation FLRW, type IV regions (clusters) are dust FLRW.}
 }
 \hfill ${}$
 \\[5mm]
 \noindent
 ${}$ \hfill
 \pb{14cm}{
 ${}$ \hfill
 \begin{tabular}{l|l|l|l|l|l|l}
 Region & Region type & Expansion rate triplet $\alpha$-$\beta$-$\gamma$ & $\rho$ & $(p_x, p_y, p_z)$ & $\Theta$ & $q$ \\
  \hline
  &&&&&\\[-3.5mm]
 I & `void' & $(0, 0, 0)$ & $0$ & $(0, 0, 0)$ & $0$ & - \\[2mm]
 IIx & $y$-$z$ `wall' & $(2/3, 0, 0)$ & $0$ & $(0, \frac{2}{9 t^2}, \frac{2}{9 t^2})$ & $\frac{2}{3 t}$ & $\frac{7}{2}$ \\[2mm]
 IIy & $z$-$x$ `wall' & $(0, 2/3, 0)$ & $0$ & $(\frac{2}{9 t^2}, 0, \frac{2}{9 t^2})$ & $\frac{2}{3 t}$ & $\frac{7}{2}$ \\[2mm]
 IIz & $x$-$y$ `wall' & $(0, 0, 2/3)$ & $0$ & $(\frac{2}{9 t^2}, \frac{2}{9 t^2}, 0)$ & $\frac{2}{3 t}$ & $\frac{7}{2}$ \\[2mm]
 IIIx & $x$ `filament' & $(0, 2/3, 2/3)$ & $\frac{4}{9 t^2}$ & $(0, \frac{2}{9 t^2}, \frac{2}{9 t^2})$ & $\frac{4}{3 t}$ & $\frac{5}{4}$ \\[2mm]
 IIIy & $y$ `filament' & $(2/3, 0, 2/3)$ & $\frac{4}{9 t^2}$ & $(\frac{2}{9 t^2}, 0, \frac{2}{9 t^2})$ & $\frac{4}{3 t}$ & $\frac{5}{4}$ \\[2mm]
 IIIz & $z$ `filament' & $(2/3, 2/3, 0)$ & $\frac{4}{9 t^2}$ & $(\frac{2}{9 t^2}, \frac{2}{9 t^2}, 0)$ & $\frac{4}{3 t}$ & $\frac{5}{4}$ \\[2mm]
 IV & `cluster' & $(2/3, 2/3, 2/3)$ & $\frac{4}{3 t^2}$ & $(0, 0, 0)$ & $\frac{2}{t}$ & $\frac{1}{2}$
 \end{tabular}
 \hfill ${}$ \\[1mm]
 {\small {\bf Table 3.}  Model (B).  Type I regions (voids) are $(0, 0, 0)$ Minkowski vacuum, type IV regions (clusters) are dust FLRW.}
 }
 \hfill ${}$
 \\[5mm]
 \noindent
 ${}$ \hfill
 \pb{14cm}{
 ${}$ \hfill
 \begin{tabular}{l|l|l|l|l|l|l}
 Region & Region type & Expansion rate triplet $\alpha$-$\beta$-$\gamma$ & $\rho$ & $(p_x, p_y, p_z)$ & $\Theta$ & $q$ \\
  \hline
  &&&&&\\[-3.5mm]
 I & `void' & $(1/10, 1/10, 1/10)$ & $\frac{3}{100 t^2}$ & $(\frac{17}{100 t^2}, \frac{17}{100 t^2}, \frac{17}{100 t^2})$ & $\frac{3}{10 t}$ & $9$ \\[2mm]
 IIx & $y$-$z$ `wall' & $(2/3, 1/10, 1/10)$ & $\frac{43}{300 t^2}$ & $(\frac{221}{900 t^2}, \frac{221}{900 t^2}, \frac{17}{100 t^2})$ & $\frac{13}{15 t}$ & $\frac{32}{13}$ \\[2mm]
 IIy & $z$-$x$ `wall' & $(1/10, 2/3, 1/10)$ & $\frac{43}{300 t^2}$ & $(\frac{221}{900 t^2}, \frac{17}{100 t^2}, \frac{221}{900 t^2})$ & $\frac{13}{15 t}$ & $\frac{32}{13}$ \\[2mm]
 IIz & $x$-$y$ `wall' & $(1/10, 1/10, 2/3)$ & $\frac{43}{300 t^2}$ & $(\frac{17}{100 t^2}, \frac{221}{900 t^2}, \frac{221}{900 t^2})$ & $\frac{13}{15 t}$ & $\frac{32}{13}$ \\[2mm]
 IIIx & $x$ `filament' & $(1/10, 2/3, 2/3)$ & $\frac{26}{45 t^2}$ & $(0, \frac{221}{900 t^2}, \frac{221}{900 t^2})$ & $\frac{43}{30 t}$ & $\frac{47}{43}$ \\[2mm]
 IIIy & $y$ `filament' & $(2/3, 1/10, 2/3)$ & $\frac{26}{45 t^2}$ & $(\frac{221}{900 t^2}, 0, \frac{221}{900 t^2})$ & $\frac{43}{30 t}$ & $\frac{47}{43}$ \\[2mm]
 IIIz & $z$ `filament' & $(2/3, 2/3, 1/10)$ & $\frac{26}{45 t^2}$ & $(\frac{221}{900 t^2}, \frac{221}{900 t^2}, 0)$ & $\frac{43}{30 t}$ & $\frac{47}{43}$ \\[2mm]
 IV & `cluster' & $(2/3, 2/3, 2/3)$ & $\frac{4}{3 t^2}$ & $(0, 0, 0)$ & $\frac{2}{t}$ & $\frac{1}{2}$
 \end{tabular}
 \hfill ${}$ \\[1mm]
 {\small {\bf Table 4.}  Model (C).  Type I regions (voids) are $(1/10, 1/10, 1/10)$ FLRW near-vacuum, type IV regions (clusters) are dust FLRW.}
 }
 \hfill ${}$
 \\[2mm]

   The behaviour of (B) \& (C) should be quite similar, and we particularly focus on (C), even though the EoS $p = (17/3) \rho$ is unrealistic, as it allows the evolution of both FLRW regions, as well as the average, to appear together on the graphs.  The evolution of model (C) is plotted in Fig.\ 2.  

 \noindent
 \pb{\textwidth}{
 ${}$ \hfill
 \includegraphics[scale=0.4]{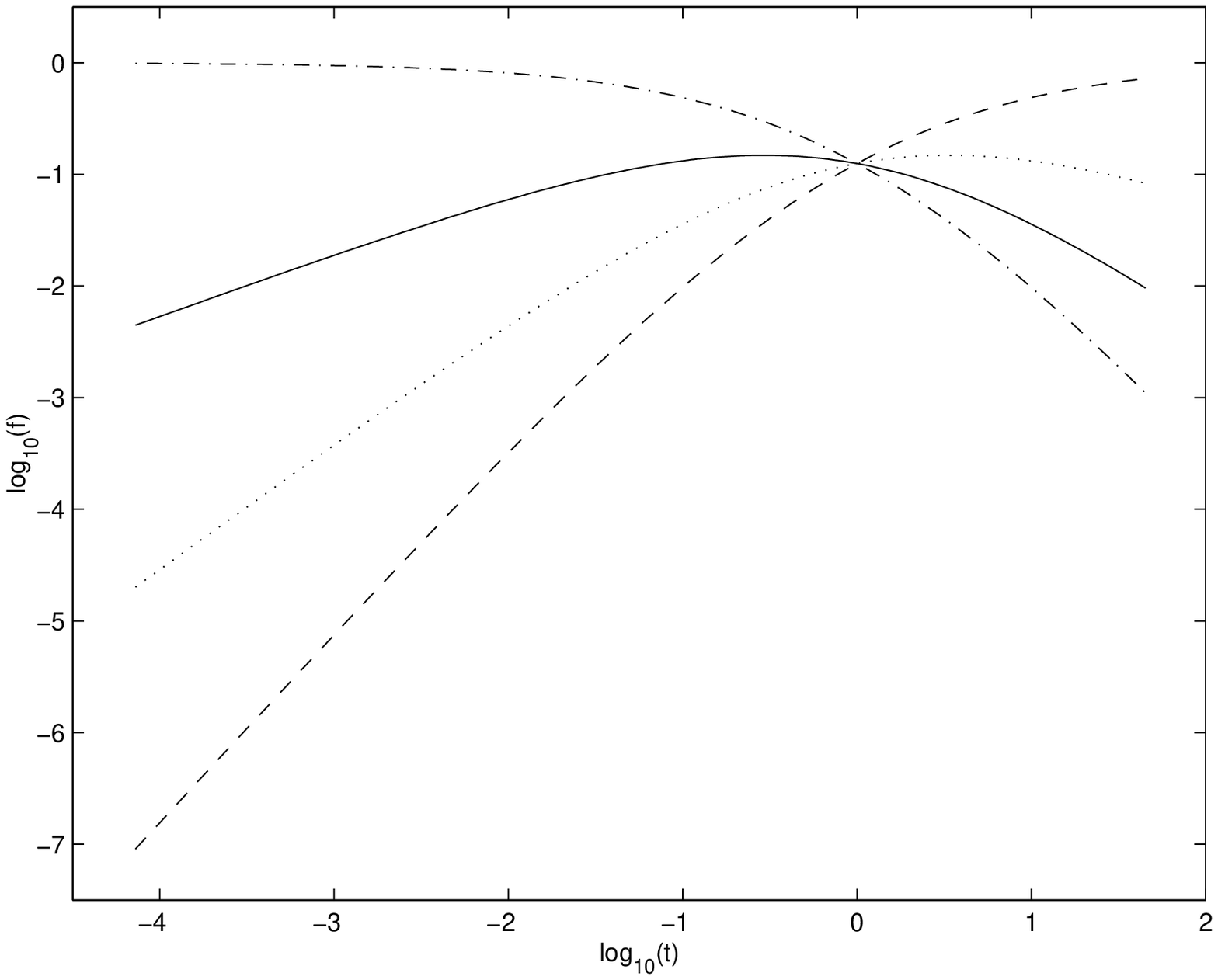}
 \hfill
 \includegraphics[scale=0.4]{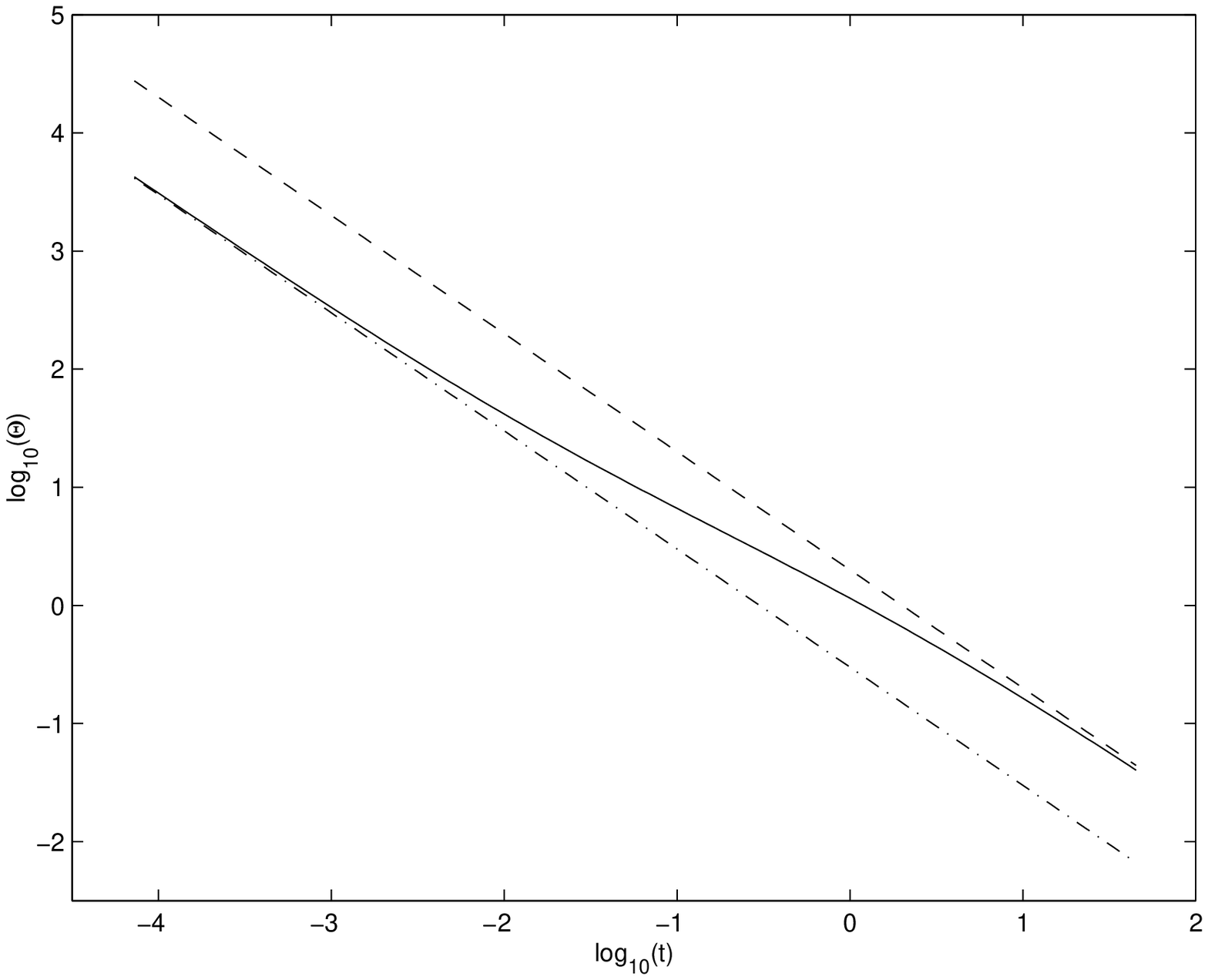}
 \hfill ${}$ \\
 ${}$ \hfill
 \includegraphics[scale=0.4]{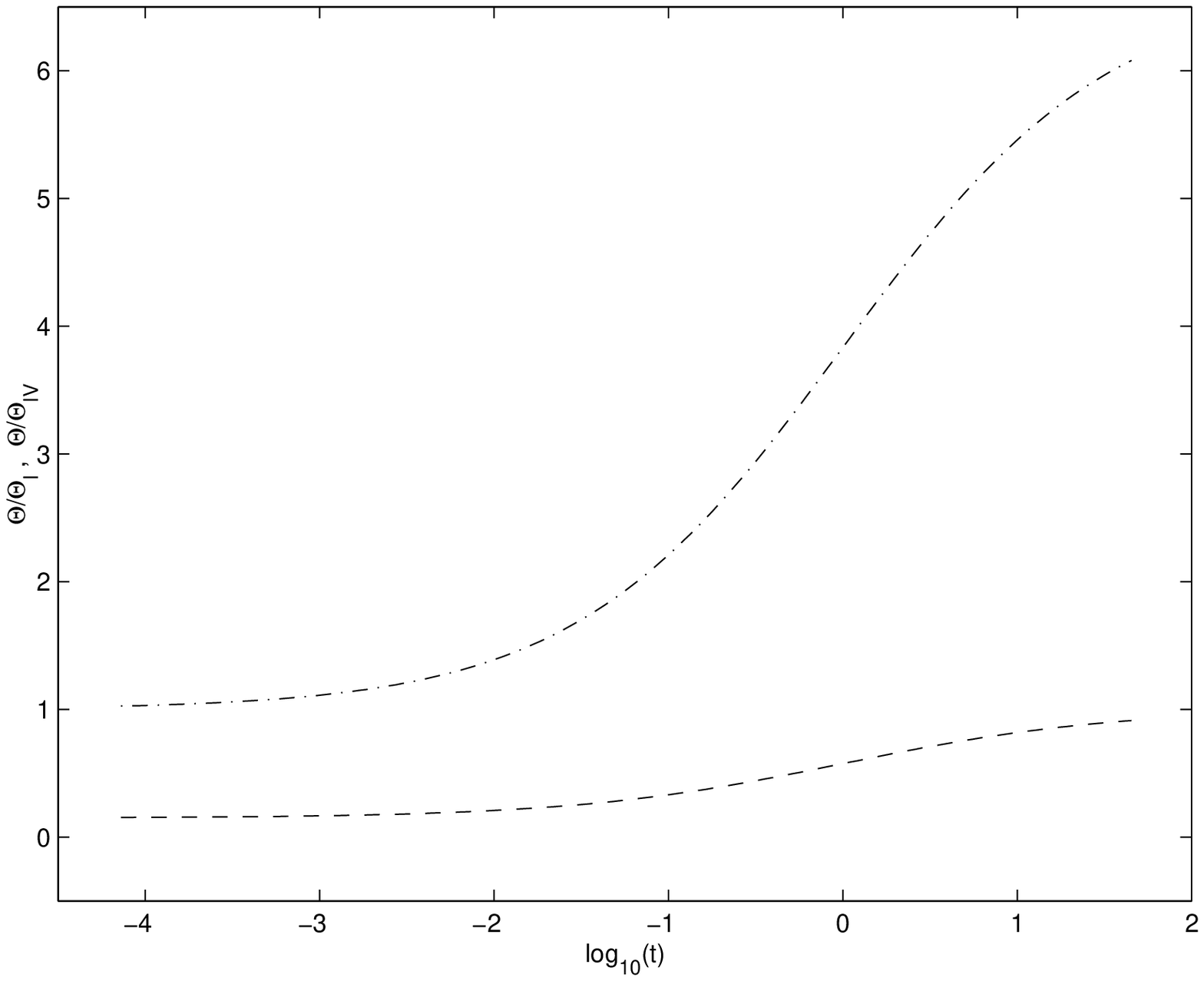}
 \hfill
 \includegraphics[scale=0.4]{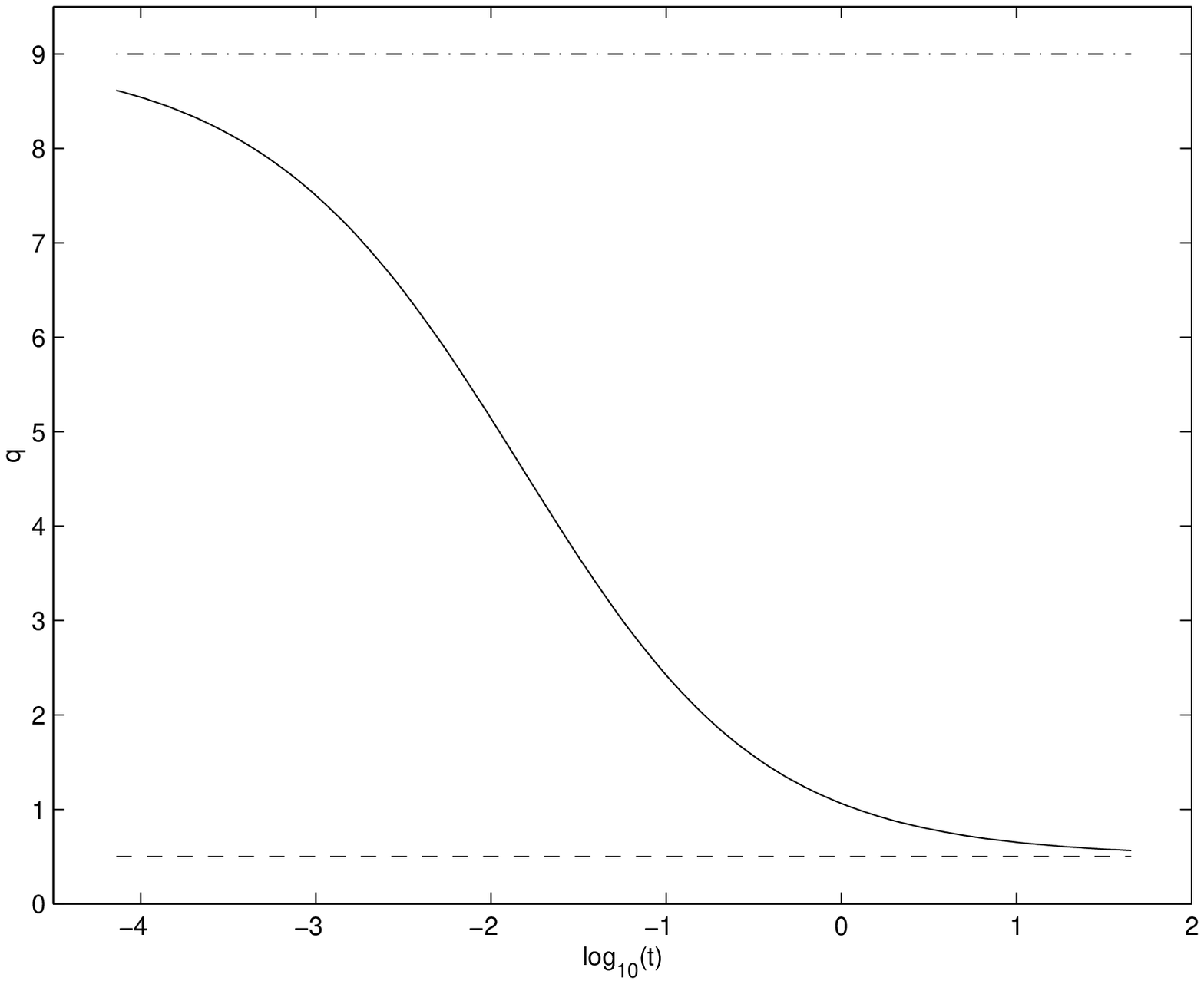}
 \hfill ${}$ \\
 ${}$ \hfill
 \includegraphics[scale=0.4]{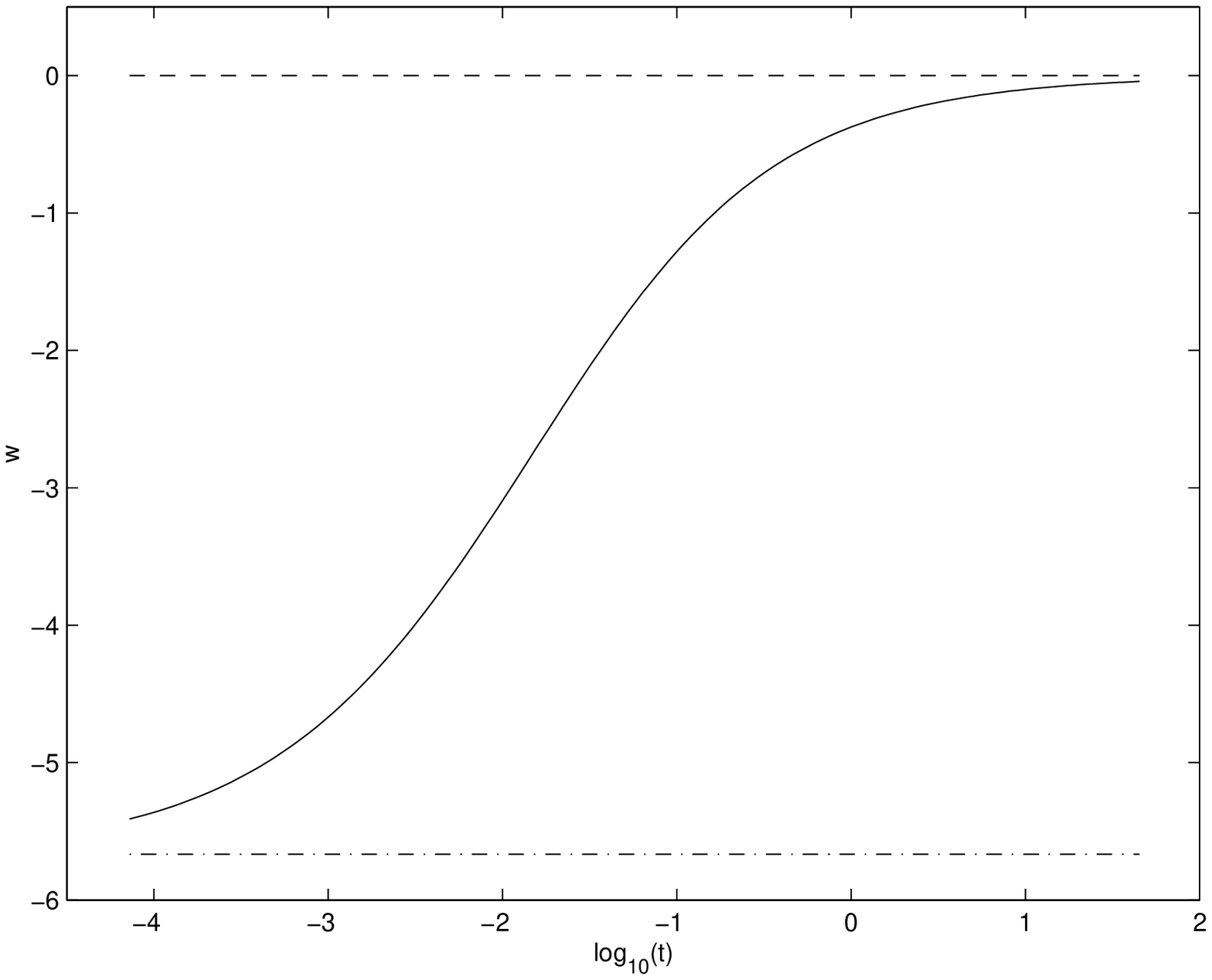}
 \hfill ${}$ \\
 ${}$ \hfill
 \pb{14cm}{
 {\small
 {\bf Figure 2.}
 Evolution of the Dust-Near-Vacuum Model, with $0.02^3 = A B C = D E F$.  Top left: evolution of the volume fraction of the 4 component types [type I (FLRW dust) - dashed, type II - dotted, type III - solid, type IV (FLRW near-vacuum) - dot-dashed]; for types II \& III there are 3 regions each with the fraction shown.  Top right: evolution of the expansion [dashed - FLRW dust, dot-dashed - FLRW near-vacuum, solid - average].  Middle left: comparison of expansion rates [dashed - $\Theta_{Av}/\Theta_I$, dot-dashed - $\Theta_{Av}/\Theta_{IV}$].  Middle right: evolution of the deceleration [dashed - FLRW dust, dot-dashed - FLRW near-vacuum, solid - average].  Bottom: evolution of the effective equation of state parameter [dashed - FLRW dust, dot-dashed - FLRW near-vacuum, solid - average].  The units are cosmological geometric units, as given in \S\ref{CGU}, for which $1$~ctu = 10~Gy, so the age of the universe is around $0.14$ on the $\log_{10} t$ axis.  The ordinate variables are dimensionless except for $\Theta$, which has units of fractional volume increase per ctu, so $\log_{10} \Theta = 1$ indicates a ten-fold increase in volume in $10$~Gy, and a Hubble rate of $65$~km/s/Mpc corresponds to $\log_{10} \Theta = 0.3$.  
 }
 }
 \hfill ${}$
 }

   Figure 2 shows the time evolution of the volume fraction of each component region, the expansion rates of the average compared with regions I \& IV, the average deceleration, and the effective EoS parameter.  This model clearly demonstrates how inhomogeneity can cause average parameters to evolve differently from what might be expected in a totally smoothed out universe.

   Although the vacuum or near-vacuum regions could model voids, the spatial flatness means they don't (or hardly) expand, whereas low density regions that develop in an initially expanding universe would retain their initial expansion --- or rather the galaxies at their borders would.  For greater realism, we would need to give the empty (type I) regions negative curvature (Milne model), and the high-density (type IV) regions positive curvature.  The type II \& III regions would need to be able to join them up across suitable surfaces and satisfy Darmois.  Thus it is important to generalise this construction method by finding matchings for other types of component region.

 \section{Discussion and Conclusions}

   A new method for constructing exact inhomogeneous cosmological models is presented.  The resulting cosmologies may have arbitrarily strong inhomogeneities on arbitrary scales, and yet they may be made very random, or exactly homogeneous on average.  They are not based on a `background' or `enveloping' metric that is effectively the large-scale average.  The method can easily be combined with the swiss-cheese approach and exact inhomogeneous metrics, and thereby it provides for a much wider range of interesting comological models, with more degrees of inhomogeneity, than hitherto.  Though the class of example models explored here is relatively simple, it does illustrate some significant features, and points the way to more complex possibilities.

   The above calculations have clearly demonstrated how the deceleration and equation-of-state parameters of the best-fit FLRW average model evolve with time, while those in each component of the construction are constant; and this even though the construction is exactly spatially uniform above a certain scale.  This behaviour is due to the component regions having different expansion behaviour.  Our finding therefore lends strong support to the contention that inhomogeneity means different regions evolve at different rates, and the evolution of the `average' model is not the just the average of those rates, but strongly depends on which regions dominate the volume.  Since we live in a region with little or no expansion, the average Riemann and Ricci fields can be very different from those felt locally.

   With the model presented here, the range of possibilities is not as large as one might like, since the spatial curvature of all the component regions is zero.  More realistically, one would expect regions of both positive and negative spatial curvature, and one would expect some regions to expand and recollapse while others are ever-expanding.  In this case, we expect the vacuum regions would expand fastest, and effective acceleration would emerge with time.  However, it is a distinctly more interesting challenge to see how a collection of such regions could be patched together, using the technique presented here.  While more general Bianchi models \cite{Bian1896,EllMcC69,RyaShe75} seem to be the obvious extension, application of the Darmois conditions between positive and negative curvature regions is likely to result in much trickier constraints.  Similarly, the limited range of constructions for this zero curvature case means that it is not possible to find completely reasonable equations of state for all the regions.  We expect generalisation of this initial model will allow for some very interesting and realistic possibilities.  Thus one should regard this current model as more illustrative than physical.

 \acknowledgments

   CH thanks David L Wiltshire \& family for excellent hosting and warm hospitality during a 7 week visit to the University of Canterbury, where long discussions were the genesis of this paper.  David should really be a co-author, but the earthquake severly curtailed his time, and he preferred to withdraw.  Thanks also to Ishwaree Neupane \& family of UC for kind hospitality.  Many thanks to the University of Canterbury for an Erskine fellowship; and the University of Cape Town for rated researcher financial support.

 \end{document}